\documentclass[%
superscriptaddress,
preprint,
showpacs,preprintnumbers,
 amsmath,amssymb,
aps,
]{revtex4-1}

\usepackage{graphicx}
\usepackage{color}
\usepackage{dcolumn}
\usepackage{bm}
\usepackage{hyperref}

\begin{document}

\title{Efficient spin injection into graphene through a tunnel barrier: overcoming the spin conductance mismatch}
\author{Qingyun Wu}
\affiliation{Department of Physics, National University of Singapore, Singapore 117542, Singapore}
\author{Lei Shen}
\email{shenlei@nus.edu.sg}
\affiliation{Engineering Science Programme, Faculty of Engineering, National University of Singapore, Singapore 117579, Singapore}
\affiliation{Department of Physics, National University of Singapore, Singapore 117542, Singapore}
\author{Zhaoqiang Bai}
\affiliation{Department of Physics, National University of Singapore, Singapore 117542, Singapore}
\author{Minggang Zeng}
\affiliation{Department of Physics, National University of Singapore, Singapore 117542, Singapore}
\author{Ming Yang}
\affiliation{Department of Physics, National University of Singapore, Singapore 117542, Singapore}
\author{Zhigao Huang}
\affiliation{College of Physics and Energy, Fujian Normal University, Fuzhou 350007, People's Republic of China}
\author{Yuan Ping Feng}
\email{phyfyp@nus.edu.sg}
\affiliation{Department of Physics, National University of Singapore, Singapore 117542, Singapore}

\date{\today}

\begin{abstract}
Employing first-principles calculations, we investigate efficiency of spin injection from a ferromagnetic (FM) electrode (Ni) into graphene and possible enhancement by using a barrier between the electrode and graphene. Three types of barriers, $h$-BN, Cu(111), and graphite, of various thickness (0-3 layers) are considered and the electrically biased conductance of the Ni/Barrier/Graphene junction are calculated. It is found that the minority spin transport channel of graphene can be strongly suppressed by the insulating $h$-BN barrier, resulting in a high spin injection efficiency. On the other hand, the calculated spin injection efficiencies of Ni/Cu/Graphene and Ni/Graphite/Graphene junctions are low, due to the spin conductance mismatch. Further examination on the electronic structure of the system reveals that the high spin injection efficiency in the presence of a tunnel barrier is due to its asymmetric effects on the two spin states of graphene.

\end{abstract}

\maketitle


\section{INTRODUCTION}
In the last decade, graphene has been the focus of intensive materials research due to its potential applications in many areas.\cite{Geim2007NM} Because of the weak spin-orbital coupling in  carbon system, graphene has long spin relaxation time and long spin diffusion length, which underlies the potential application of graphene in spintronics. A key step for realizing graphene-based spintronic devices is injection of a spin current into graphene and which has been the focus of many studies. The first work on spin transport in graphene was reported in 2006 by Hill \textit{et~al.}\cite{Hill2006ITM} Using ferromagnetic (FM) NiFe electrodes, they observed a spin valve effect, in which a spin-polarized current injected from one ferromagnetic electrode, goes through graphene before being detected at the other electrode. This idea was soon pursued by several other groups.\cite{Han2009APL,Tombros2007Nature, Popinciuc2009PRB, Goto2008APL} However, the reported spin injection efficiency, a key parameter for spin transport, was only about 1~\% if graphene is directly
in contact with the FM leads.\cite{Hill2006ITM,Han2009APL,Goto2008APL} The low spin injection efficiency is mainly due to mismatch of spin ``conductance" \cite{Schmidt2000PRB,Rashba2000PRB,Fert2001PRB} between the ferromagnet and graphene. In a typical heterostructure of a ferromagnetic electrode and a nonmagnetic (NM) material such
as graphene, spins injected from the electrode into the NM material may diffuse through the NM material, or back scattered to the lead.\cite{Schmidt2000PRB,Rashba2000PRB,Fert2001PRB} The flow of the spin current via diffusion depends on the spin resistance of the NM material as well as matching of the resistances of the two materials at the FM/NM interface.\cite{Takahashi2003PRB} The reported ratio of spin resistance of ferromagnet ($R_{FM}$) over that of graphene ($R_G$) varies from $10^{-3}$ to $10^{-5}$.\cite{Han2010PRL} Because $R_{FM}\ll R_G$, spin diffusion in a typical FM/graphene junction is dominated by the back scattering of spins into the FM lead, which is the reason for the low spin injection efficiency.\cite{Han2010PRL} To enhance the spin injection efficiency, insulating oxides such as Al$_2$O$_3$ and MgO were used as a tunnel barrier between the ferromagnetic electrode and graphene \cite{Dlubak2010APL, Tombros2007Nature, Jozsa2008PRL, Jozsa2009PRB, Jozsa2009PRB2, Popinciuc2009PRB}. It has been demonstrated that if the interfacial spin-dependent resistivity can be tuned to the same order of magnitude as the spin-dependent resistivity of graphene,  efficient spin injection can be achieved. \cite{Goto2008APL}
However, not all systems exhibit perfect barrier effects. Using Al$_2$O$_3$ and
MgO as tunnel barrier, respectively, Tombros {\em et al.}\cite{Tombros2007Nature} and
Han {\em et al.}\cite{Han2009APL} separately reported good match of contact resistances
but low spin injection efficiency from Co electrode to graphene, which was attributed
to issues related the materials and their interface. For example, pinholes were observed
in the Al$_2$O$_3$ barrier by van Wees' group which may create short circuit between
graphene and the ferromagnetic electrode, resulting in a reduced spin injection
efficiency  \cite{Tombros2007Nature, Popinciuc2009PRB}.
To reduce pinholes, Dlubak and co-workers recently proposed to use sputtering technique
for sample growth\cite{Dlubak2012APL}. On the other hand, formation of clumps of the
insulating material on graphene was reported by Kawakami's group which also leads to
low spin injection efficiency. This is due in part to graphene's reluctance to form
strong bonds with other materials, and can be overcome by incorporating an interfacial
TiO$_2$ layer \cite{Han2010PRL}. Nevertheless, identify materials with good conductance
match and thus high spin injection efficiency remains a challenge for graphene-based spintronic applications.

Motivated by recent experimental breakthroughs in synthesizing $h$-BN/Graphene heterostructure and the vertical Graphene/$h$-BN/Graphene field-effect-transistor (FET),\cite{Britnell2012Science,Roth2013NL,Haigh2012NM} here we investigate spin injection from a Ni electrode into graphene with an $h$-BN tunnel barrier, as well as Cu and graphite barriers for comparison. Results of our transport calculations indicate that the insulating $h$-BN tunnel barrier significantly improves the spin injection efficiency from the Ni lead into graphene. This is possible because of an $h$-BN induced asymmetry in the spin states of graphene, as revealed by our first-principles electronic structure calculations.

\section{MODEL}

Graphene was first isolated by mechanical exfoliation from graphite. This method was later applied to other layer-structured materials, creating a family of two-dimensional materials that includes insulating $h$-BN, and semiconducting MoS$_2$ and WS$_2$,
in addition to the semi-metallic graphene. These materials can easily form van der Waals
structures with well-defined interfacial contact.\cite{Britnell2012NL, Britnell2012Science, Georgiou2013NN, Choi2013NC, Yu2012NM, Haigh2012NM, Geim2013Nature} In particular,
$h$-BN has been demonstrated to be the best candidate for graphene-based heterostructures or sandwich structures because of its wide band gap and close match of its lattice constant with that of graphene and ferromagnetic Ni.\cite{Britnell2012Science,Geim2013Nature,Liu2012PRB,Maassen2011NL,Karpan2008PRB,Karpan2007PRL} Good carrier transport and tunneling properties of the graphene/$h$-BN heterostructure or sandwich structures have been reported by Geim's group.\cite{Haigh2012NM,Britnell2012Science,Georgiou2013NN,Geim2013Nature} Furthermore, previous first-principles calculations predicted that spin-polarized tunneling can be achieved in Ni(111)/$h$-BN \cite{Yazyev2009PRB,Karpan2011PRB,Giovannetti2007PRB} and Ni(111)/graphene heterostructures \cite{Liu2012PRB,Maassen2011NL,Karpan2008PRB,Karpan2007PRL}, as well as Ni(111)/graphene/Ni(111) spin-valve devices.\cite{Cho2011JPCC,Saha2012PRB}
Therefore, $h$-BN is selected as a tunnel barrier for spin injection from Ni to graphene.
For comparison, we also consider a metallic Cu barrier, as well as a thin graphite layer
or a few-layer graphene barrier.
Ni(111) is selected as the ferromagnetic source because it has a similar hexagonal lattice
structure as graphene and $h$-BN, and its in-plane lattice constant (2.49 \AA) is also
close to that of graphene (2.46 \AA) and $h$-BN (2.50~\AA). The lattice constant of
graphene was adopted for the in-plane lattice constant of the Ni(111)/$h$-BN/Grpahene
structure. The small stain induced in Ni(111) and $h$-BN is not expected to make any qualitative difference in the calculated results.

The model used in our study is shown in \textbf{Fig.~1}. Both the left lead (Ni) and the
right lead (graphene) are assumed infinitely long. We also consider a case of Ni/($h$-BN)$_3$/graphene with doubled overlapping area between Ni and BN as well as BN and graphene. We found a little change of the spin injection efficiency. Thus, we use the structure in Fig.~1 in this work. The scattering region or the device consists of the Ni lead at the bottom, the graphene active region at the top, and 0-3
layers of $h$-BN (or Cu, graphene) between them. For convenience of discussion, we assume
the transport direction is along the $y$-direction and the normal direction of the graphene
plane is the $z$-direction. The system is periodic along the $x$-direction, as shown in
\textbf{Fig.~1}.
It is noted that a single
piece of graphene is used for the active region and the right lead which would minimize
interface scattering between the central region and the right lead in the usual sandwich
structures.

A separate slab model, consisting a monolayer graphene, 6 layers of Ni, and 0 to 3
layers of $h$-BN was used for structural optimization and band structure calculation.
During geometry optimization, the bottom four layers of Ni were fixed to their positions
in bulk Ni while all other
atoms in the system were allowed to freely relax. In the optimized structure, the distance between the graphene layer and the Ni substrate is 2.04~{\AA}, while the BN-BN and
BN-graphene interlayer distances  are 3.25~\AA~and 3.44~\AA, respectively (see \textbf{Fig.1}), which are in good agreement with results of previous theoretical
studies.\cite{Avramov2012JAP} The BN layer was found to interact strongly with Ni(111).
In the optimized structure, the N atom in $h$-BN sits on the top of the Ni atom in the
surface layer, forming a N-Ni bond of length 2.06~\AA. The B atom in $h$-BN is
directly above the Ni atom in the third layer, as shown in \textbf{Fig.~1}. The A-A stacking
is assumed between BN and graphene because it is the most stable configuration among various possible stackings. The model for transport calculation was constructed from the optimized slab model.

\section{COMPUTATIONAL DETAILS}
Geometry optimization and electronic structure calculation were carried
out using the density functional theory (DFT) based \textsc{VASP} code.\cite{Kresse1996PRB,Kresse1996CMS} The projector augmented wave (PAW) method and the local density approximation (LDA)\cite{Perdew1981PRB} were adopted to describe the core-valence interactions and the electron exchange and correlation functional. The plane-wave expansion of electron wavefunction was cutoff at a kinetic energy of 400 eV and the Brillouin zone of the unit cell was sampled using a 21$\times$21$\times$1~$k$-point grid. Structural optimization was carried out until the force on each atom is less than 0.01~eV/{\AA}. The transport properties such as $I$-$V$ curve were studied using a self-consistent approach that combines DFT and the non-equilibrium Green's function (NEGF) formalism, as implemented in the \textsc{ATK} package,\cite{Brandbyge2002PRB,Taylor2001PRB} 
in which the electron transmission used in the Landauer's formula is obtained self-consistently for a given value of bias voltage, same as in other previous works.\cite{Saha2012PRB,Cho2011JPCC,Shen2012PRB,Shen2012PRB2,Bai2013PRB} The double-$\zeta$ polarized (DZP) basis set was used to expand the electron wave function in transport
calculation and a cutoff energy of 150 Ry and a Monkhorst-Pack $k$-point grid of $9\times 1\times 100$ yielded a good balance between accuracy and computational time. The LDA was also adopted in the transport calculations and the electron temperature was set to 300 K. A finer $k$ point mesh ($201\times 1$) was used to sample the periodic direction perpendicular to the transport direction. A vacuum region of 15~{\AA} was used to separate the device from its periodic image to minimize the artificial interaction between them.

\section{RESULTS AND DISCUSSIONS}
\subsection{Transport property}
\subsubsection{Ni(111)/graphene junction without tunnel barrier}
Before we examine the effect of a tunnel barrier, we calculate and present here the
transport property of Ni(111)/graphene junction in which graphene is directly in contact
with the Ni(111) electrode. The device structure is shown in \textbf{Fig.~2a}. It is
similar to the structure shown in \textbf{Fig.~1} but without the $h$-BN layers.
The calculated transmission spectrum of the Ni (111)/graphene
junction under a bias voltage of 0.3~V is show in \textbf{Fig.~2c} where the bias voltage window is
within the two vertical dashed lines. As can be seen, the total transmissions of spin up
and spin down states are similar within the bias voltage window, except a peak near -0.10~eV
in the spin down state. This implies that the spin polarization of the current passing
through the device will not be high if the Ni lead is directly deposited on graphene.

For a more qualitative measure of spin polarization, {\em i.e.}, spin injection efficiency
from Ni lead into graphene, we calculate the $I$-$V$ curves using the NEGF approach.
The spin resolved current $I_{\sigma}$, where $\sigma$ indicates the spin up or spin
down state, is obtained from
\begin{equation}
I_{\sigma}=\frac{e}{h}\int_{-\infty}^{+\infty}T_{\sigma}(E,V)[f_{L}(E,\mu
_{L})-f_{R}(E,\mu_{R})]dE,
\end{equation}
where $e$, $h$, and $T$ are the electron charge, Planck's constant, and the transmission,
respectively. $f_L$ and $f_R$ in the above equation are the Fermi distribution functions
of the left and right lead, respectively. Under a bias voltage $V$, the chemical
potentials of the left lead and right lead are shifted to $\mu_{L}=E_{F}-eV/2$ and $\mu_{R}=E_{F}+eV/2$, respectively. The calculated $I$-$V$ curve of the Ni(111)/graphene
junction when the bias voltage is varied from 0 to 0.3~V is shown in \textbf{Fig.~2b}.
It is clear that spin down electrons are the majority spin in the current. The spin
down current is larger than the spin up current, but not overwhelmingly. This transport
property can be partially explained by the electronic structure at the Fermi surface
of Ni(111) and graphene.\cite{Karpan2007PRL} When the Fermi surface of {\em fcc} Ni
and graphene are projected to the (111) plane, higher density of states is found for
down spin near the six high-symmetry points (\textbf{K} or \textbf{K}') which are the
main transport channels of graphene. In contrast, the Fermi surface states for up
spin of Ni are located elsewhere.\cite{Karpan2007PRL} Therefore, spin down electrons
dominate over spin up electrons in the current under a bias voltage (\textbf{Fig~.2b}).
The spin injection efficiency is calculated from $(I_{\rm up}-I_{\rm down}) / (I_{\rm up}+I_{\rm down})$. Under a bias voltage of 0.3~V, the estimated spin injection efficiency of Ni(111)/graphene is 48\%. This, however, is an optimistic value estimated based on an ideal model. Under experimental conditions, this value is expected
to be reduced by interfacial effect such as interfacial disorder.\cite{Tombros2007Nature,Popinciuc2009PRB,Jozsa2009PRB,Voloshina2011NJP,Zhang2013APA,Zhang2012APL}

\subsubsection{Ni(111)/graphene junction with $h$-BN tunnel barrier}

Spin tunneling has been proposed as a way of overcoming the conductance mismatch and has
been widely used to enhance the spin injection efficiency of spintronic devices, such as silicon-based devices.\cite{Jansen2012NM} It is therefore natural to ask whether the same
approach works here and can be used to enhance spin injection efficiency from a FM lead
to graphene. To investigate the effect of a tunnel barrier,
we insert \textit{n} ($n=1,2,3$) layers of $h$-BN between Ni(111) and monolayer graphene, as illustrated in \textbf{Fig.~1}. The model used in our transport calculation is also shown
in \textbf{Fig.~3a}.
The spin tunneling transport properties of the Ni(111)/$h$-BN/graphene were calculated
and the results are presented in \textbf{Fig.~3}. As shown in \textbf{Fig.~3c}, the transmission spectrum of spin up states under a bias voltage of 0.3~V is greatly
suppressed within the bias voltage window while that of the spin down states remains significant.
The calculated $I$-$V$ curve of the Ni(111)/$h$-BN/graphene device with three layers of
$h$-BN (\textbf{Fig.~3b}) confirms that the majority spin current is much larger than the
minority spin current under a bias voltage.
The estimated spin injection efficiency of the Ni(111)/$h$-BN/graphene device with 1-3
layers of $h$-BN are given in \textbf{Table 1}. The results indicate that the spin injection efficiency from the ferromagnet into graphene can be dramatically enhanced by tunneling
through an $h$-BN barrier. With a single layer of $h$-BN, the spin injection efficiency
increased to 72\%, which is close to the experimental results of 64\% \cite{Yamaguchi2013arxiv}, and much higher than 1\% of FM/graphene \cite{Han2009APL} or 30\%
of FM/Oxide/graphene.\cite{Han2010PRL} The maximum spin injection efficiency (100\%)
can be achieved with three or more layers of $h$-BN between the FM lead and graphene. The enhancement of the spin injection efficiency is due to the improvement of conductance mismatch. Figure 2b and Figure 3b show the current ($I$) is reduced from $\mu$A to $n$A after inserting $h$-BN. It means that the resistance of FM electrodes is increased by $h$-BN barriers around 3 order of magnitude. Thus the resistance of R$_F$ and R$_G$ is close, indicating a good conductance match.

To understand how the $h$-BN tunnel barrier affects the spin up and spin down currents, we
calculated the spin-resolved transmission eigenstates and present the results for the
device with three layers of $h$-BN in \textbf{Fig.~4}. For the spin up states, we can see
clearly in \textbf{Fig. 4a} that most of the transmission states are localized in the
first two layers of $h$-BN from the Ni electrode and a little on the N atoms in the third
$h$-BN layer which is next to graphene. The eigenstates in the third $h$-BN layer show
clear $p$ characteristics. Therefore, the transport channel of spin up states of graphene
is essentially blocked. On the other hand, the spin down transmission states are
delocalized in all $h$-BN layers as well as graphene, as show in \textbf{Fig.~4b}.
The transport channel of the spin down states of graphene is thus open, and the spin
down electrons can be easily injected from the Ni electrode to graphene. It is noted that the transport channel in graphene is of carbon $p_z$ characteristic (\textbf{Fig.~4b}).
Based on the spin-resolved transmission eigenstates of the two transport channels, we
present a schematic diagram in \textbf{Fig.~4c} to illustrate the effect of the insulating
barrier on the electron transport property of the proposed structure.
Because of improved conductance matching by using a tunnel barrier,\cite{Han2010PRL} spin polarized electric current is
injected into graphene from a ferromagnetic electrode. The injected electric current is highly spin polarized because the majority spin transport channel
is blocked by the tunnel barrier.

\subsubsection{Ni(111)/graphene junction with Cu or graphite barrier}
Ni(111)/graphene/Ni(111) junction was proposed to have large magetoresistance (MR).\cite{Karpan2008PRB, Karpan2007PRL, Saha2012PRB,Giovannetti2008PRL} The MR ratio (pessimistic) can reach 100\% if five or more layers of graphene are used.
But if a monolayer graphene is sandwiched between the open $d$-shell transition-metals,
such as Ni, its characteristic electronic structure of topological singularities at the
$K$-points in the reciprocal space would be destroyed by the formation of strong chemical bonds between graphene and transitional-metal electrodes, leading to a low MR ratio.
Karpan {\em et al.}, proposed to insert several layers of inert Cu to avoid bond
formation between graphene and metal lead.\cite{Karpan2008PRB} It was found that with a single layer of Cu
between Ni and graphene, the electronic structure of graphene can be restored and its MR
ratio can reach 90\%. The MR ratio can be further increased by incorporating more layers
of graphene between metal leads.\cite{Karpan2008PRB}

Of course, magnetoresistance is different from spin injection ratio. The former relies on
the magnetic configuration of the two electrodes, and is an extrinsic property of a
system, while spin injection ratio is determined by the spin-dependent behaviors of
injected electrons and it is an intrinsic property of the system.
Despite of this difference, it is interesting to find out whether a Cu or graphite barrier is useful for
enhancing the spin injection efficiency between graphene and the Ni(111) electrode. With this in mind and also to serve as a comparison with the $h$-BN tunnel barrier, we also calculated the spin injection efficiencies of Ni(111)/graphene junctions with a few atomic layers of Cu (111) or graphite as barrier and list the results in \textbf{Table~1}.
It is clear that the spin injection ratio is low with either graphite or Cu (111) as
the barrier. The reason is because a few layers of graphite or Cu are metallic. The
lack of tunneling effect makes the metallic barriers less effective in overcoming the
conductance mismatch between Ni and graphene, compared to an insulating barrier such as
$h$-BN. Experimentally, it was demonstrated that
a thin titanium seed layer between TM-electrodes and graphene improves the contact
conductivity and lattice match, but does not lead to enhancement in spin injection ratio,\cite{Han2010PRL} due to the same reason. Nevertheless, based on the results of our calculations, a good
spin injection ratio (79\%) may be achieved if a monolayer Cu (111) is inserted between
the Ni (111) electrode and graphene. This is because the Cu layer weakens the interaction
between Ni and graphene, allowing graphene to recover its characteristic electronic
property. However, since Cu is metallic, electron transport through Cu is not by    tunneling. If the thickness of the Cu barrier is more than one atomic layer, the spin injection ratio
is hampered by the conductance mismatch between Cu and graphene. By comparing the
performances of the three different barriers (\textbf{Table 1}), it is obvious that the insulating $h$-BN, which interacts with graphene through the van der Waals force, is the
most promising barrier to facilitate spin injection from a FM
electrode into graphene, for graphene-based spintronic applications.

\subsection{Electronic structures }
\subsubsection{Band structures}
From the results of our transport calculations presented above, we know that the $h$-BN
tunnel barrier is effective in promoting spin injection from a TM-electrode into graphene.
In order to understand the underlying physics, we calculated the spin-resolved band
structure and local density of states of the Ni(111)/$h$-BN/graphene structures.
The results obtained for the structure with a single layer of $h$-BN between Ni (111) and graphene are shown in \textbf{Fig.~6}, along with those without a $h$-BN barrier for
comparison. The solid circles in \textbf{Fig.~6} represent the weight of the graphene-derived $p_{z}$ orbital. We pay attention to the graphene-derived $p_{z}$ orbital because it is the main transport channel as shown in \textbf{Fig.~4b}.
As shown in \textbf{Figs.~6a and 6b}, when the Ni electrode is in direct contact with
graphene, a band gap of about 0.34 eV opens in both spin up and spin down bands of graphene, which is in agreement with results of previous experiments and calculations.\cite{Karpan2008PRB, Zhou2007NM, Giovannetti2007PRB} The gap opening is due to strong interaction between graphene and Ni. The similar band structures for majority and minority
carriers implies a low spin injection efficiency if the ferromagnetic Ni lead is directly
deposit on graphene.
In fact, because of the conductance mismatch between Ni and graphene ($R_{\rm Ni}/R_{\rm graphene}$ ranges from $10^{-3}$ to $10^{-5}$)\cite{Han2010PRL}, most of the charge carriers would be backscattered to Ni at the Ni/grpahene interface.
Moreover, the measured spin polarization can be further reduced by
interfacial effect and/or interfacial disorder in experimentally grown
samples.\cite{Ke2008PRL,Ke2010PRL} All these result in a much lower spin injection efficiency into graphene.\cite{Tombros2007Nature,Popinciuc2009PRB,Jozsa2009PRB}
Interestingly, when a layer of $h$-BN is incorporated between Ni and graphene, a band gap
of about 85~meV opens in the spin up bands of graphene, while the spin down bands
remain gap less, as shown in \textbf{Figs.~6c and 6d}. In other words, the semi-metallic characteristics of graphene is restored
in the spin down states by the $h$-BN layer. It is this $h$-BN induced imbalance between
the two spin states of graphene that results in the significantly different transport performance of the two spin channels. This is the root of the high spin injection efficiency of the Ni(111)/$h$-BN/grpahene device.

\subsubsection{Local density of states}
To understand why an $h$-BN layer induces asymmetric effects on the two spin states of
graphene, we examine the local density of states (LDOS) projected on the relevant atoms
and orbitals. As can be seen in \textbf{Fig.~7a}, there exists a strong overlap between
the C-$p$ and Ni-$d$ orbitals in both spin up states, in the energy range of -0.23 to -0.3 eV, and spin down states, between 0.23 and 0.28 eV, if metallic Ni is in direct contact with graphene.
When one atomic layer of $h$-BN is incorporated between Ni and graphene, the interaction
between the C-$p$ and Ni-$d$ orbitals is eliminated in the spin down states, as shown in \textbf{Fig.~7b}. However, a weak coupling between these orbitals still exists in the
spin up states, in the energy range of -0.3 to -0.32 eV. Further examination of the PDOS
of other atoms reveals that this weak coupling is mediated by the N-$p_z$ orbitals
(not shown here). These features in the PDOS of the concerned atoms and close relationship
between DOS and transmission lend further support that Ni/$h$-BN/graphene is an efficient
tunneling barrier interface.

\section{CONCLUSIONS}
The issue of spin conductance mismatch and resulting low spin injection efficiency which hampers the practical application of graphene in spintronics is addressed using first-principles electronic structure and transport calculations. $h$-BN was found to be an effective tunnel barrier for enhancing the spin current injection efficiency from ferromagnetic electrodes into graphene. Our study suggests that tunneling transport can efficiently overcome the spin conductance mismatch between ferromagnetic electrodes
and graphene, similar to other channel materials. Recently, Yamaguchi {\em et~al.} demonstrated spin injection into bilayer graphene from ferromagnetic Ni$_{0.8}$Fe$_{0.2}$ electrodes through a single-crystal monolayer $h$-BN.\cite{Yamaguchi2013arxiv} These studies pave the way for spintronic devices based on graphene as well as other 2D materials.
Note that two-terminal devices should require more considerations besides the spin injection efficiency but will need to be treated along the guidelines in ref.7-9.

\section{Acknowledgments}
Authors thank S. C. Li, P. J. Kelly for their helpful comments on the discussion of Cu(111) and graphite barriers and asymmetric characteristics of two spin states of graphene with the $h$-BN barrier. This work is partially supported by A*STAR (Singapore) Research Funding (Grant No. 092-156-0121).

\clearpage
\begin{table*}
\caption{\label{tab:table1}The calculated spin injection efficiency of Ni(111)/barrier/Graphene under a bias voltage of 0.3 V. Three different barriers, $h$-BN, graphene, and Cu (111), of different thickness (1, 2 or 3 atomic layers) are considered.}
\centering
\begin{tabular}{l|cccc}
\hline
\hline
Barrier & \multicolumn{3}{c}{Spin injection efficiency} \\
  & 1L & 2L & 3L \\
\hline
  $h$-BN    & 72\% & 96\% & 100\% \\
  Graphene  & 29\% & 31\% & 24\%  \\
  Cu (111)  & 79\% & 12\% & 13\%  \\
\hline
\hline
\end{tabular}
\end{table*}

\clearpage

\begin{figure}
\centering
\includegraphics[width=0.8\textwidth]{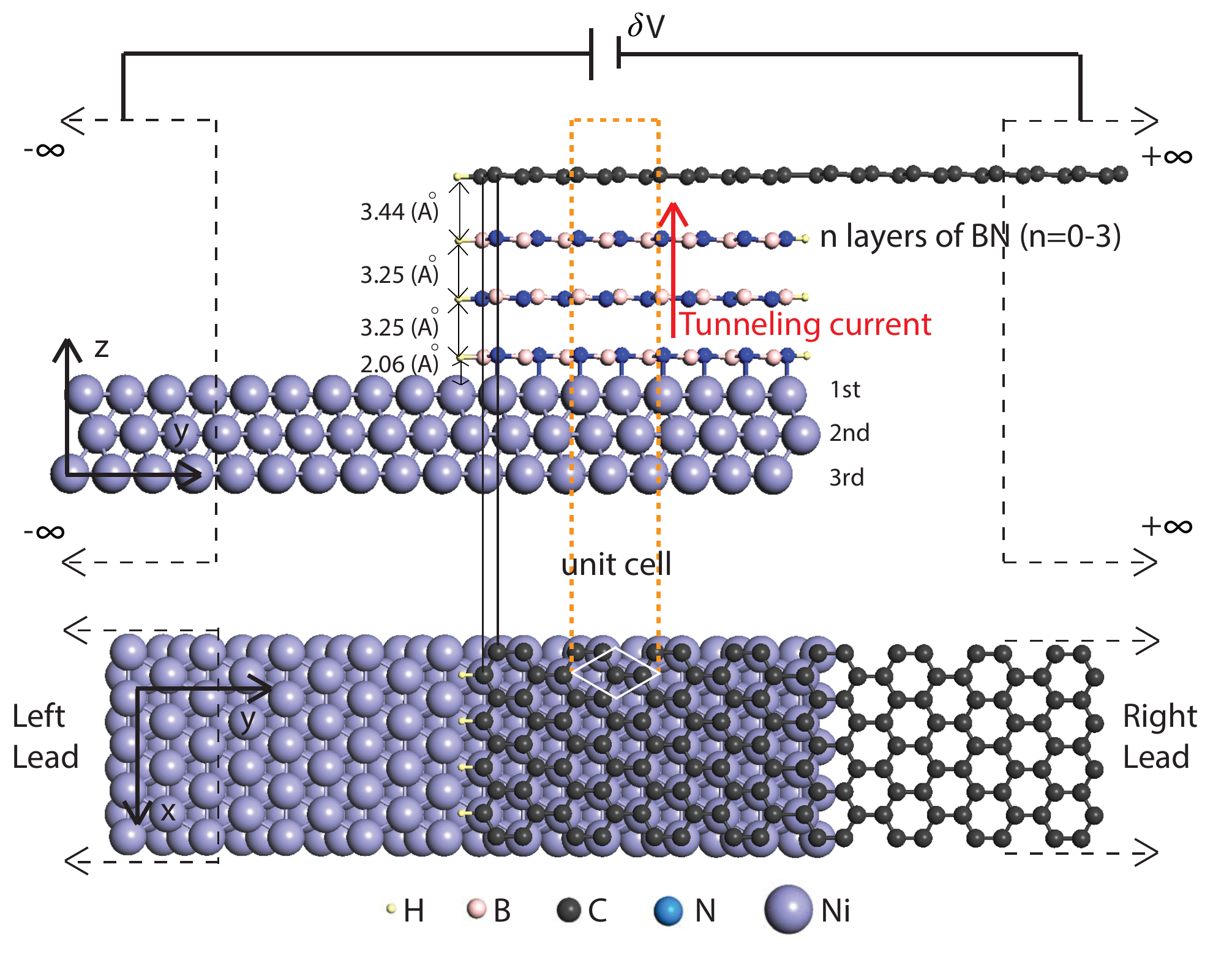}\\
\caption{Side and top views of the Ni(111)/$h$-BN/Graphene model. The thickness of the $h$-BN tunnel layer (or Cu, graphite) is varied from 0 (without tunnel barrier) to 3 atomic layers. The
model shown here has 3 atomic layers of $h$-BN. The supercell used for structural
optimization and band structure calculation is indicated by the brown dashed box.
The device is built with two semi-infinite leads and a scattering region.}
\end{figure}

\begin{figure}
\centering
\includegraphics[width=0.8\textwidth]{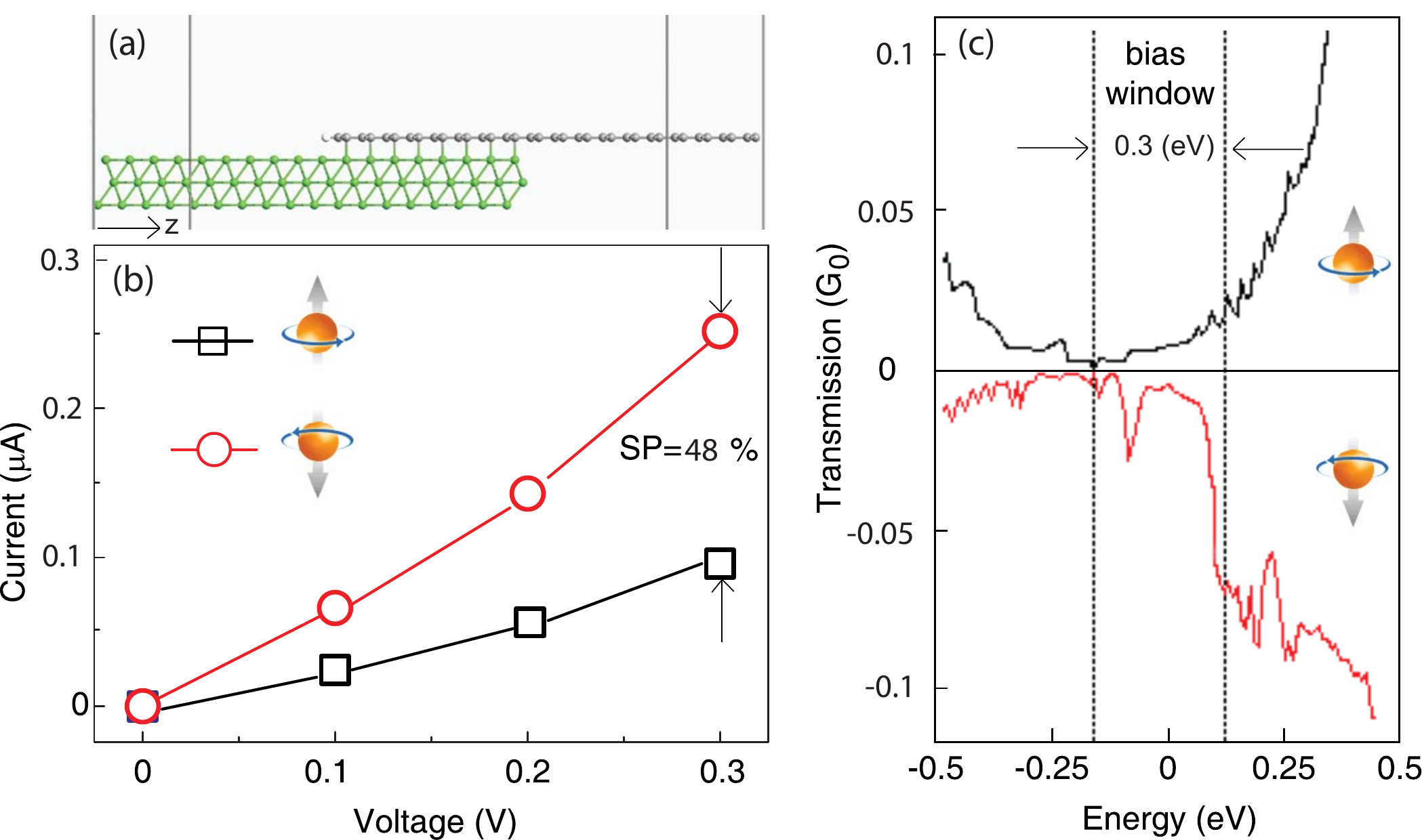}\\
\caption{(a) Model for Ni(111)/graphene. (b) The calculated $I$-$V$ curve of Ni(111)/graphene. (c) The transmission spectrum of Ni(111)/graphene under a bias voltage of 0.3~V.}
\end{figure}

\begin{figure}
\centering
\includegraphics[width=0.8\textwidth]{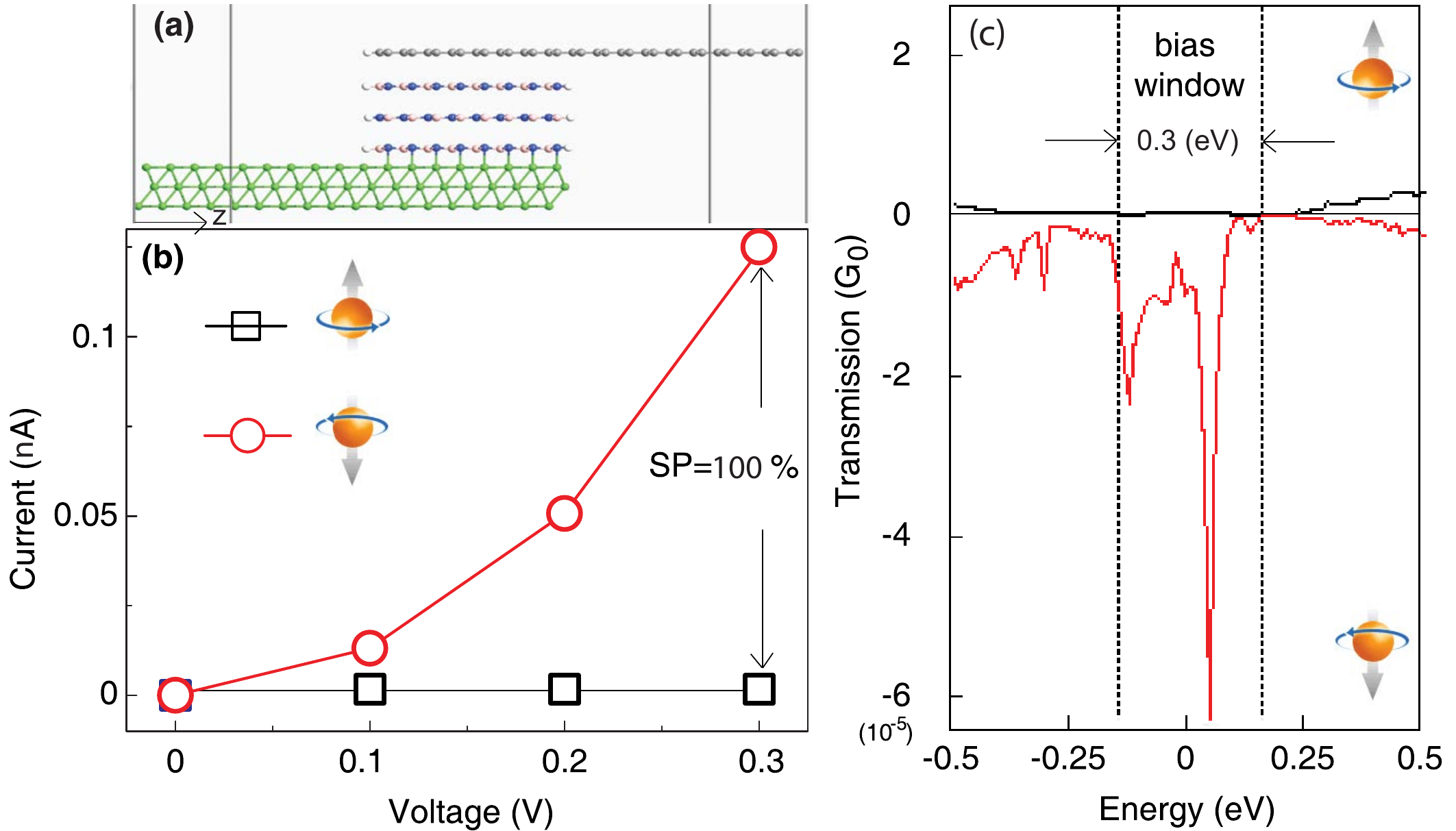}\\
\caption{(a) Model for Ni(111)/($h$-BN)$_3$/graphene. The subscript 3 indicates 3 atomic layers of $h$-BN. (b) The calculated $I$-$V$ curve of Ni(111)/($h$-BN)$_3$/graphene. (c) The transmission spectrum of Ni(111)/($h$-BN)$_3$/graphene under a bias voltage of 0.3~V.}
\end{figure}

\begin{figure}
\centering
\includegraphics[width=0.8\textwidth]{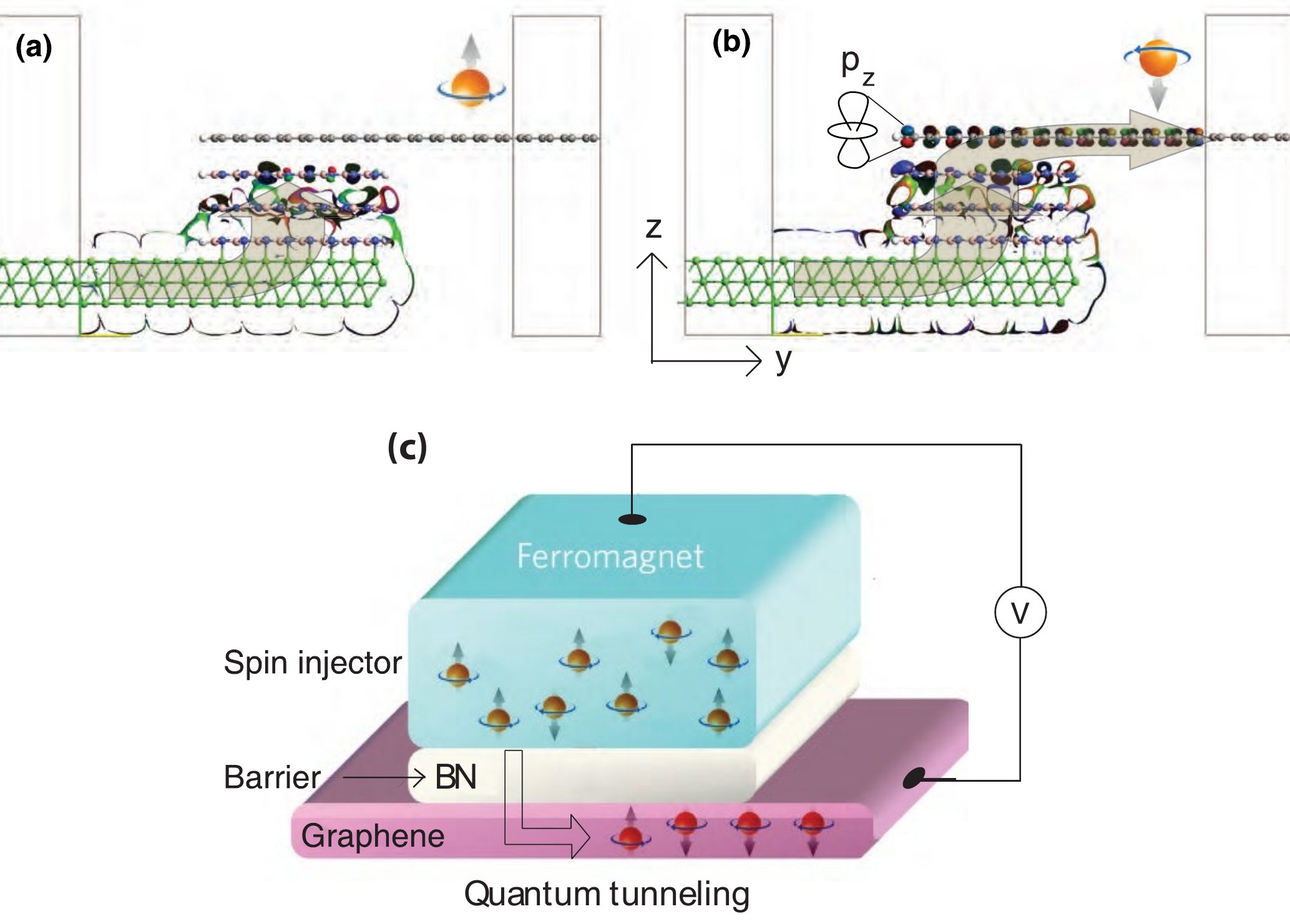}\\
\caption{The calculated spin-resolved transmission eigenstates of  (a) the spin up channel and (b) spin down channel of Ni(111)/($h$-BN)$_3$/graphene under a bias voltage of 0.3~V. (c) Schematic diagram showing effect of a tunnel barrier on spin polarization of current
injected from the Ni(111) electrode to graphene.}
\end{figure}

\begin{figure}
\centering
\includegraphics[width=0.8\textwidth]{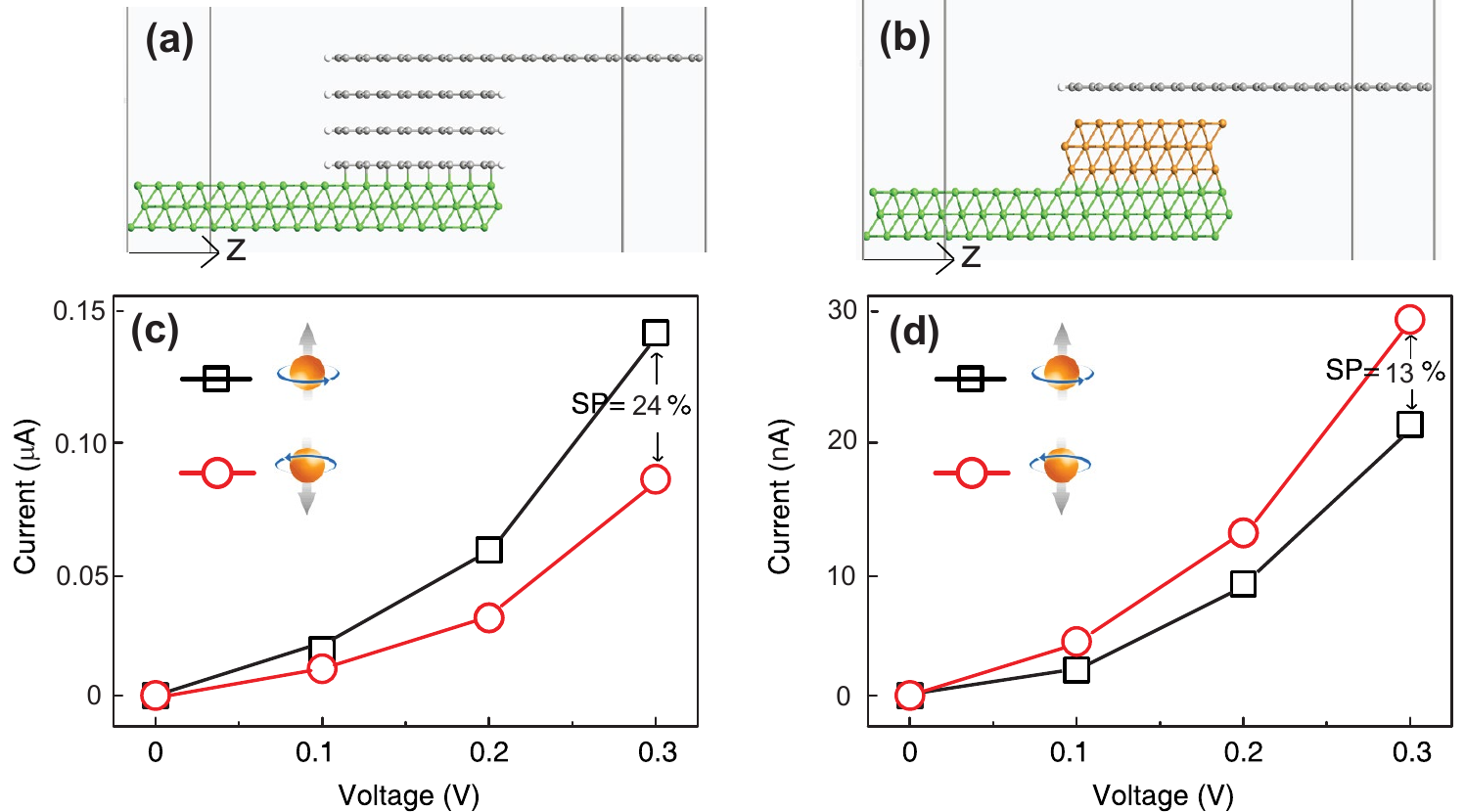}\\
\caption{Models for (a) Ni(111)/(graphite)$_3$/graphene and (b) Ni(111)/(Cu)$_3$/graphene. The subscript indicates the
thickness of the barrier in number of atomic layers. (c) and (d) The calculated $I$-$V$ curves of the above two devices, respectively. }
\end{figure}

\begin{figure}
\centering
\includegraphics[width=0.8\textwidth]{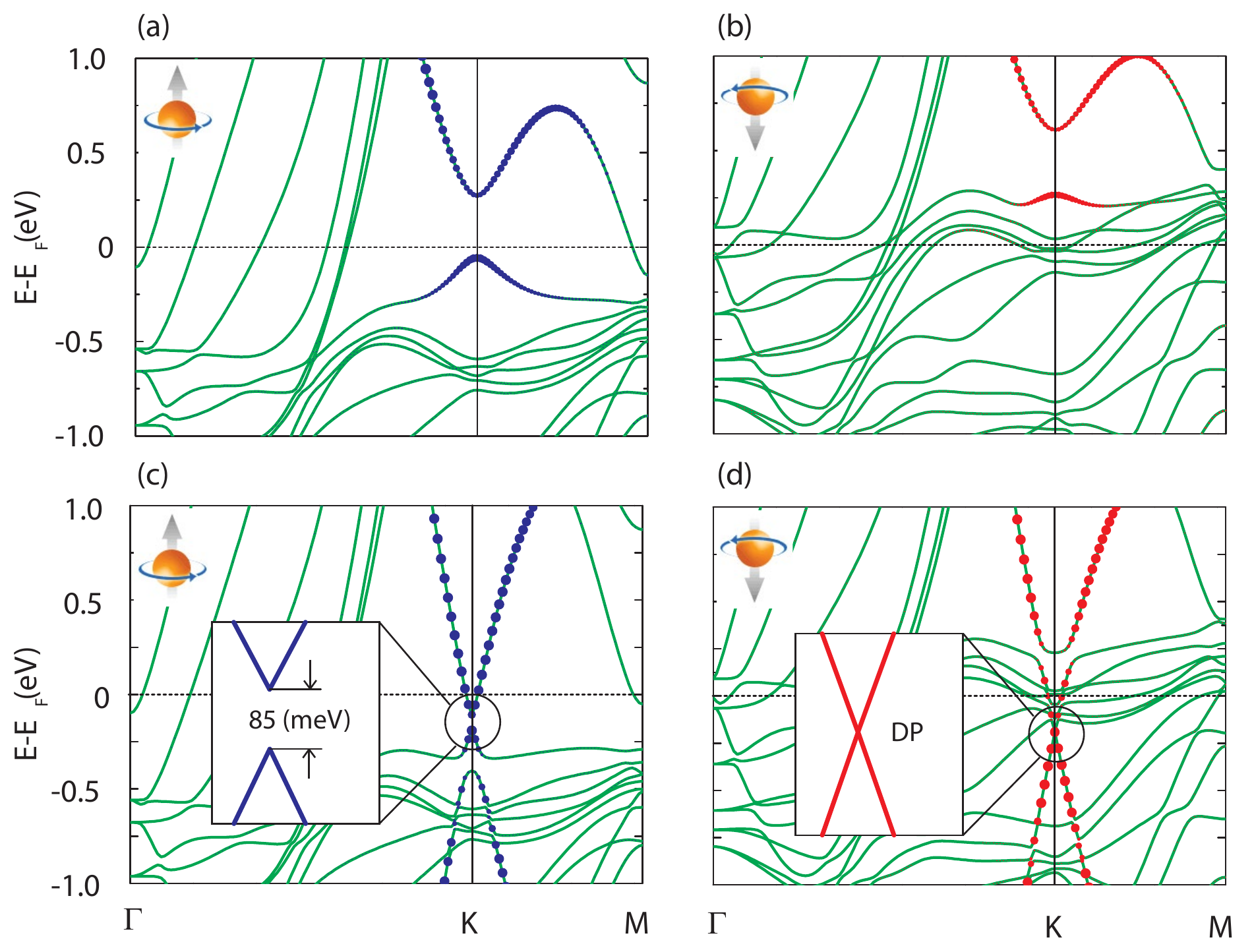}\\
\caption{Spin resolved band structures of Ni(111)/graphene without a tunnel barrier (upper panels) and with one atomic layer of $h$-BN (lower panels), respectively. The spin up (down) band structure is shown in the left (right) panel, in each case. The solid circles represent   the weight of the graphene-derived $p_z$ orbital.}
\end{figure}

\begin{figure}
\centering
\includegraphics[width=0.6\textwidth]{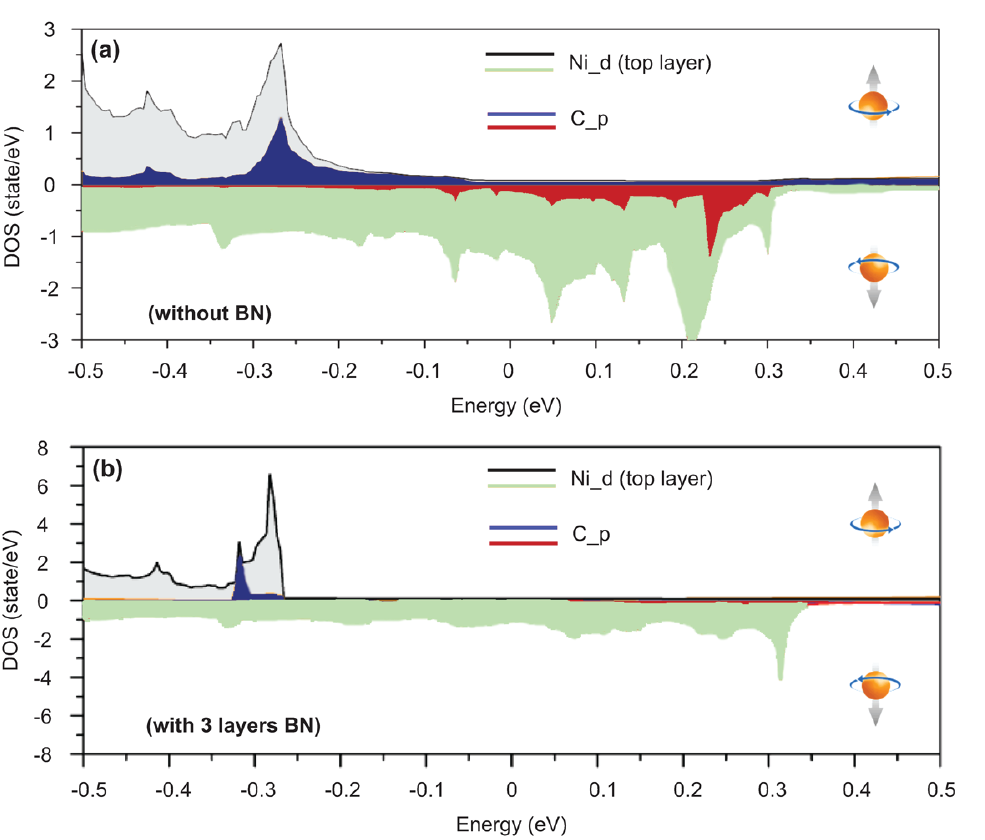}\\
\caption{LDOS of Ni (111)/graphene (a) without a tunnel barrier and (b) with one atomic layer of $h$-BN, respectively. The LDOS on B and N orbitals are not shown. To show the details, the LDOS of C-$p$ are enlarged by 5 times. As can be seen, the overlap of spin down states of C-$p$ and Ni-$d$ disappears after inserting $h$-BN.}
\end{figure}

\end{document}